\documentclass{article}
\usepackage{ltexpprt}

\usepackage{times}
\usepackage[english]{babel}

\usepackage{algorithm}
\usepackage{algorithmic}

\usepackage{amsmath}
\usepackage{amssymb}
\usepackage{array}
\usepackage{booktabs}
\usepackage[small]{caption}
\usepackage{cite}
\usepackage{color}
\usepackage{dsfont}
\usepackage{epsfig}
\usepackage{fmtcount}
\usepackage{footnote}
\usepackage{graphicx}
\usepackage{hyperref}
\usepackage{mdwlist}
\usepackage{microtype}
\usepackage{multirow}
\usepackage{placeins}
\usepackage{rotating}
\usepackage{subcaption}
\usepackage{tikz}
\usetikzlibrary{decorations.pathreplacing}
\usepackage{url}
\usepackage{verbatim}
\usepackage{xspace}
\usepackage{xcolor}
\usepackage{enumitem}

\usepackage{amsfonts}

\usepackage{color}
\newcommand{\LN}{\ensuremath{L_\mathbb{N}}}
\newcommand{\LU}[2]{\ensuremath{\log \binom{#1}{#2}}}

\newcommand{\mi}[1]{\mathit{#1}}
\newcommand{\mb}[1]{\mathbf{#1}}
\newcommand{\mc}[1]{\mathcal{#1}}
\def\A{${\mathbf A}$}
\def\E{${\mathbf E}$}
\def\M{${\mathbf M}$}
\def\model{${\mathit M}$}

\newtheorem{observation}{Observation}

\newtheorem{problem}{Problem}
\newtheorem{example}{Example}

\newcommand{\mdl}{L}

\newcommand{\hide}[1]{}
\newcommand{\reminder}[1]{{\textsf{\textcolor{red}{[#1]}}}}

\newcommand{\slashburn}{\textsc{SlashBurn}\xspace}

\newcommand{\method}{\textsc{VoG}\xspace}
\newcommand{\original}{\textsc{Original}\xspace}
\newcommand{\methodsb}{\textsc{VoG}\xspace}

\newcommand{\gnf}{{\textsc{Greedy'nForget}}\xspace}

\newcommand{\tophun}{{\textsc{Top100}}\xspace}
\newcommand{\topten}{{\textsc{Top10}}\xspace}
\newcommand{\topk}{{\textsc{Top-k}}\xspace}
\newcommand{\all}{{\textsc{Plain}}\xspace}
\newcommand{\plain}{{\textsc{Plain}}\xspace}

\newcommand{\ben}{\begin{enumerate*}}
\newcommand{\een}{\end{enumerate*}}
\newcommand{\bit}{\begin{itemize*}}
\newcommand{\eit}{\end{itemize*}}

\newcommand{\Model}{M}
\newcommand{\Models}{\mathcal{M}}
\newcommand{\DB}{\mathcal{D}}

\newcommand{\G}{G}

\newcommand{\problemFormulation}{{\bf Problem Formulation:}\xspace}
\newcommand{\scalability}{{\bf Effective and Scalable Algorithm:}\xspace}
\newcommand{\experiments}{{\bf Experiments on Real Graphs:}\xspace}

\newcommand{\Flickr}{{\tt Flickr}\xspace}
\newcommand{\WWWBarabasi}{{\tt WWW-Barabasi}\xspace}
\newcommand{\Epinions}{{\tt Epinions}\xspace}
\newcommand{\Enron}{{\tt Enron}\xspace}
\newcommand{\ASOregon}{{\tt AS-Oregon}\xspace}
\newcommand{\Wikipediachoc}{{\tt Chocolate}\xspace}
\newcommand{\Wikipediacontro}{{\tt Controversy}\xspace}
\newcommand{\Cavemen}{{\tt Cavemen}\xspace}

\newif\ifapx
\apxtrue
\apxfalse

\begin{document}

\clubpenalty=10000
\widowpenalty = 10000

\title{{\method}: Summarizing and Understanding Large Graphs}

\author{
 Danai Koutra\\
{\small\hspace*{-1em}School of Computer Science\hspace*{-1em}}\\
{\small Carnegie Mellon University}\\
{\small {danai@cs.cmu.edu}}
 \and 
U Kang\\
{\small\hspace*{-1.2em}Computer Science Department\hspace*{-0.8em}}\\
{\small KAIST}\\
{\small ukang@cs.kaist.ac.kr}
 \and
Jilles Vreeken\\
{\small\hspace*{-0.8em}Max Planck Institute for Informatics\hspace*{-1.2em}}\\
{\small and Saarland University}\\
{\small jilles@mpi-inf.mpg.de}
 \and 
Christos Faloutsos\\
{\small\hspace{-1em}School of Computer Science\hspace*{-1em}}\\
{\small Carnegie Mellon University}\\
{\small christos@cs.cmu.edu}
}
\date{}

\maketitle

\begin{abstract}
How can we succinctly describe a million-node graph with a few simple sentences? 
How can we measure the `importance' of a set of discovered subgraphs in a large graph?
These are exactly the problems we focus on. 
Our main ideas are to construct a `vocabulary' of
subgraph-types that often occur in real graphs (e.g., stars, cliques, chains), and from a set of subgraphs, find the most succinct description of a graph in terms of this vocabulary. We measure success in a well-founded way by means of the Minimum Description Length (MDL) principle: a subgraph is included in the summary if it decreases the total description length of the graph.

Our contributions are three-fold:
(a) {\em formulation}: we provide a principled encoding scheme to choose
vocabulary subgraphs;
(b) {\em algorithm}: we develop \method, an efficient method
to minimize the description cost,
and
(c) {\em applicability}:
we report experimental results on multi-million-edge real graphs,
including Flickr and the Notre Dame web graph.

%

\end{abstract}


\section{Introduction}
\label{sec:intro}

Given a large graph, say, a social network like Facebook,
what can we say about its structure?
As most real graphs, the edge distribution
will likely follow a power law~\cite{DBLP:conf/sigcomm/FaloutsosFF99},
but apart from that, is it random?
If not, how can we efficiently and in simple terms summarize which parts of the graph stand out, and how?
%
%
%
The focus of this paper is exactly finding short summaries for large graphs, in order to gain a better understanding of their characteristics.

Why not apply one of the many
community detection, clustering or graph-cut algorithms
that abound in the literature \cite{Chakrabarti:2004, DBLP:conf/www/LeskovecLDM08,PrakashPAKDD2010,Karypis99@METIS,cook:94:subdue}, and summarize the graph in terms of its communities? The answer is that these algorithms do not quite serve our goal. Typically they detect numerous communities without explicit ordering, so a principled selection procedure of the most ``important'' subgraphs is still needed. In addition to that, these methods merely return the discovered communities, without characterizing them (e.g., clique, star), and, thus, do not help the user gain further insights in the properties of the graph. 

In this paper, we propose
\method, an efficient and
effective method to summarize and understand
large real-world graphs, in particular graphs beyond the so-called ``cavemen'' networks that only consist of well-defined, tightly-knit clusters (cliques or near-cliques).


The first insight is to best {\em describe} the structures in a graph using
an enriched set of ``vocabulary'' terms: cliques and near-cliques (which are typically considered by community detection methods), and also stars, chains  and (near) bi-partite cores.
The reasons we have chosen these ``vocabulary'' terms are:
(a) (near-) cliques are included, and so our method works fine
on ``cavemen'' graphs, and
(b) stars \cite{KangF11}, chains \cite{TauroPSF01} and bi-partite cores \cite{kleinberg99web,PrakashPAKDD2010} appear very often, and have semantic meaning (e.g., factions, bots)
in the tens of real networks we have seen in practice
(e.g., IMDB movie-actor graph, co-authorship networks,
netflix movie recommendations,
US Patent dataset, phonecall networks).

The second insight is to {\em formalize} our goal
as a lossless compression problem,
and use the MDL principle.
 The best summary of a graph is the set of subgraphs
that describes the graph most succinctly, i.e.,
compresses it best, and, thus, helps a human understand the main graph characteristics in a simple, non-redundant manner. 
A big advantage is that our approach is \emph{parameter-free}, as at any stage MDL identifies the best choice: the one by which we save most bits.

Informally, we tackle the following problem: 
\vspace{-0.7em}
\begin{problem}[Informal]
\label{prob:informal}
\hfill \bit
\vspace{-0.7em}
\item {\bf Given}{\emph :} a graph 
\item {\bf Find}{\emph :} a set of possibly overlapping subgraphs
\item \textbf{to most succinctly describe} the given graph, i.e., explain as many of its edges in as simple possible terms,
\item in a {\bf scalable} way, ideally 
linear on the number of edges.
\eit
\end{problem}



\begin{figure*}[t!]
        \centering
        \begin{subfigure}[tb]{0.23\textwidth} 
               \centering
               \includegraphics[width=0.58\textwidth]{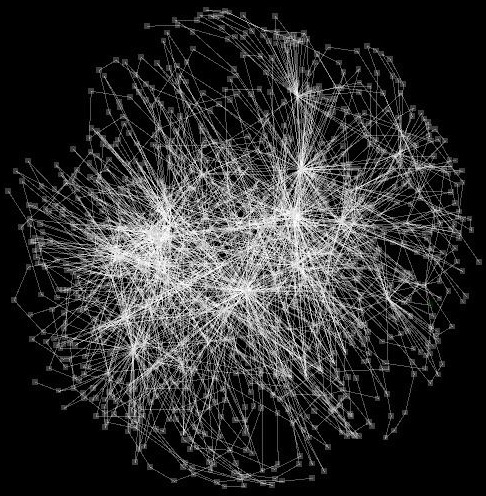}
               \caption{Original Wikipedia \Wikipediacontro graph
	       (with `spring embedded' layout \cite{SpringModel}). No structure stands out.
	       \hfill  }
               \label{fig:lcrOrig}
        \end{subfigure}
        ~
	 \begin{subfigure}[tb]{0.1cm}
         \begin{tikzpicture}[scale=1.0]
             \draw [thick] (0,0) -- (0,4.5);
         \end{tikzpicture}
	 \end{subfigure}
        ~
        \begin{subfigure}[tb]{0.23\textwidth} 
               \centering
               \includegraphics[width=0.58\textwidth]{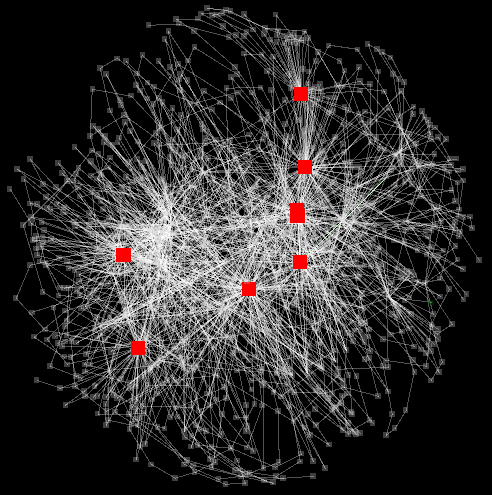}
               \caption{\method: 8 out of the 10 most informative
	       structures are stars
	       (their centers in red - Wikipedia editors,
	       heavy contributors etc.).  \label{fig:lcrStars}}
        \end{subfigure}
        ~ %
        \begin{subfigure}[tb]{0.23\textwidth} 
                \centering
                \includegraphics[width=0.65\textwidth]{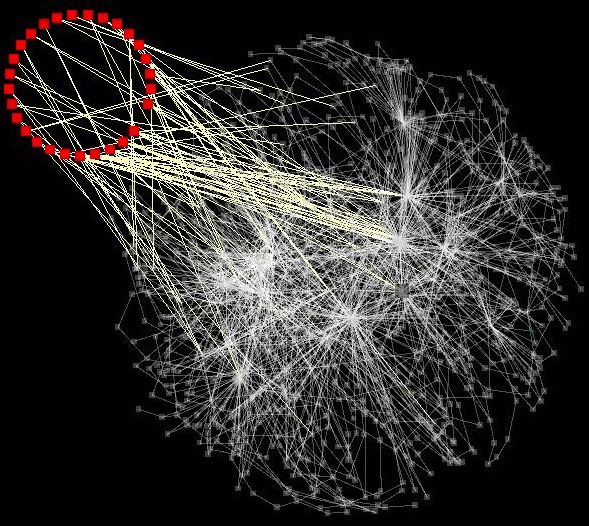}
                \caption{\method: The most informative bipartite graph - `edit war' -
		warring factions (one of them, in the top-left red circle), changing each-other's edits.}
                \label{fig:lcrTop1nb}
        \end{subfigure}
        ~ 
        \begin{subfigure}[tb]{0.23\textwidth} 
                \centering
                \includegraphics[width=0.68\textwidth]{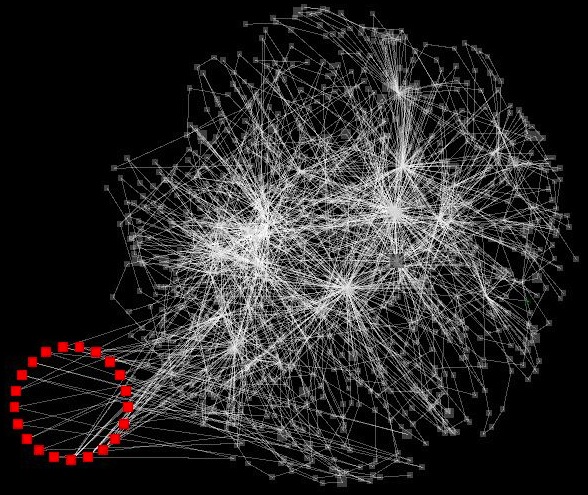}
                \caption{\method: the second most informative bipartite graph  - another
		`edit war', between vandals
		(bottom left circle of red points) vs
		responsible editors (in white).}
                \label{fig:lcrTop2nb}
        \end{subfigure}
        \caption{\method: summarization and understanding of the most informative, from an information theoretic point of view, structures of the Wikipedia \Wikipediacontro
	graph.
	Nodes stand for  Wikipedia contributors and
	edges link users who edited the same part of the article.
	Without \method, in \ref{fig:lcrOrig}, no clear structures stand out.
	\method
	spots stars in \ref{fig:lcrStars} (Wikipedia editors and other
	heavy contributors),
	and bipartite graphs in
	\ref{fig:lcrTop1nb} and \ref{fig:lcrTop2nb} (reflecting
	`edit wars', i.e., editors reverting others' edits).
        Specifically, \ref{fig:lcrTop1nb} shows
	the dispute between the two parties
	about a controversial topic
	and
	\ref{fig:lcrTop2nb} shows
	vandals (red circles) vs responsible
	Wikipedia editors. }
        \label{fig:lcrMediaWiki}
\end{figure*}
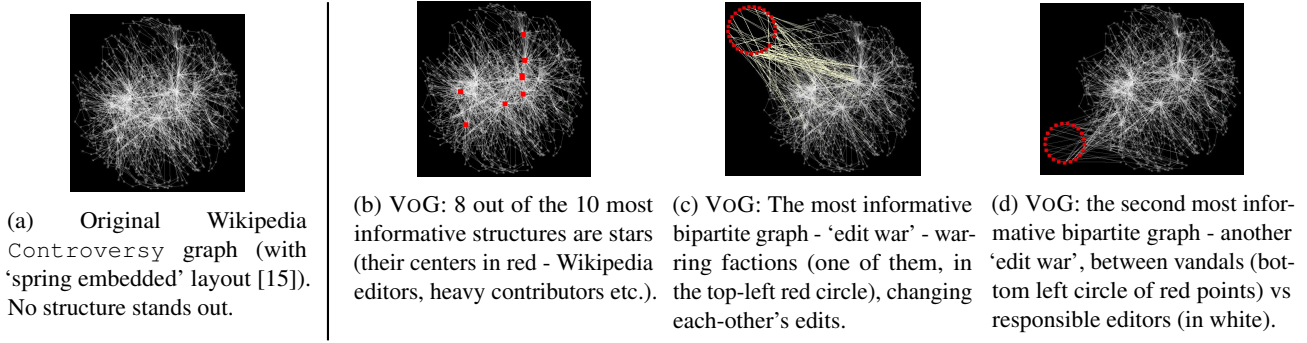

\hide{
    There are two main insights in our approach:

    The first is to {\em formalize} our goal,
    as a lossless compression problem
    (good compression means we discovered significant regularities in the graph).
    The second insight is to carefully choose the right primitives
    (`graph vocabulary': blocks, stars, chains, etc.),
    such that we can summarize a graph
    in easily understood terms
    We iterate, that, if the input graph is a ``caveman'' graph,
    then \method will automatically choose
    only blocks, to achieve a good compression.
}


\noindent and our contributions can be summarized as:
\begin{enumerate*}
  \item \problemFormulation
      We show how to formalize the intuitive concept of
      graph understanding
      using principled, information theoretic arguments.
  \item \scalability
      We design \method 
      which is near-linear on the number of edges.
  \item \experiments
     We empirically evaluate \method
      on several real, public graphs 
      spanning up to millions of edges.
      \method spots interesting patterns
      like `edit wars' in the Wikipedia graphs (Fig.~\ref{fig:lcrMediaWiki}).
\end{enumerate*}

\enlargethispage{-\baselineskip}
The paper outline is standard:
overview, problem formulation, method description,
experiments, and conclusions.
Due to lack of space, we give more details and experiments in  
the Appendix. 

\section{Proposed Method: Overview and Motivation}
\label{sec:overview}
Before we give our two main contributions in the next sections -- the problem formulation, and the search algorithm --,
we first provide the high-level outline of \method,  
%
which stands
for {\em Vocabulary-based summarization of Graphs}:
\vspace{-0.2cm}
\begin{itemize*}
\item[(a)] We use MDL to formulate a quality function:
a collection ${M}$ of structures
(e.g., a star here, cliques there, etc) is as good as its description length $L(G, M)$. Hence, any subgraph or set of subgraphs has a quality score.
\item[(b)] We give an efficient algorithm for characterizing candidate subgraphs.
In fact, we allow {\em any} subgraph discovery heuristic to be used for this, as we define our framework in general terms and use MDL to identify the structure \emph{type} of the candidates.
\item[(c)] Given a candidate set ${\cal C}$ of promising subgraphs,
we show how to mine informative summaries, removing redundancy  by minimizing the cost.
\end{itemize*}
\vspace{-0.15cm}
\method results in a list $M$ of, possibly overlapping subgraphs, sorted in importance order (compression gain). Together these succinctly describe the main connectivity of the graph.


The motivation behind \method is that people cannot easily understand cluttered graphs, whereas a handful of simple structures are easily understood, and often meaningful. Next we give an illustrating example of \method, where the most `important' vocabulary subgraphs that constitute a Wikipedia article's (graph) summary are semantically interesting.

{\em Illustrating Example:}
In Fig.~\ref{fig:lcrMediaWiki}
we give the results of \method on the Wikipedia \Wikipediacontro graph;
the nodes are editors, and
editors share an edge if they edited the same part of the article.
Figure~\ref{fig:lcrOrig} shows the graph using
the spring-embedded model~\cite{SpringModel}.
No clear pattern emerges, and thus a human would have hard time
understanding this graph. 
Contrast that with the results of \method. 
Figs.~\ref{fig:lcrStars}--\ref{fig:lcrTop2nb} depict the same graph,
where we highlight the most important structures
(i.e., structures that save the most bits) discovered by \method.
\vspace{-0.15cm}
\begin{itemize*}
\item {\em Stars $\rightarrow$ admins (+ vandals)}:
in Fig.~\ref{fig:lcrStars}, with red color, we show
the centers of the most important ``stars'':
further inspection shows
that these centers
typically correspond to administrators 
who revert vandalisms and make corrections. 
%
\item {\em Bipartite cores $\rightarrow$ edit wars}:
Figs.~\ref{fig:lcrTop1nb} and \ref{fig:lcrTop2nb}
give the two most important near-bipartite-cores.
Manual inspection shows that these correspond to {\em edit wars}:
two groups of editors reverting each others' changes. For clarity, we denote the members of one group by red nodes (left), 
and hi-light the edges to the other group in pale yellow.
\end{itemize*}

\section{Problem Formulation}
\label{sec:encoding}

\hide{
	In the next two sections we give our two major contributions:
	the problem formulation, and the search algorithm.
	First, however, we give
	the high-level outline of \method, which stands
	for {\em Vocabulary-based summarization of Graphs}:
	\begin{enumerate}
	\item[(a)] We use MDL to formulate a quality function:
	a collection ${M}$ of structures
	(e.g., a star here, cliques there, etc) is as good as its description length $L(G, M)$. Any subgraph (set) hence has a quality score.
	\item[(b)] We give an efficient algorithm for extracting candidate subgraphs.
	In fact, we allow {\em any} subgraph discovery heuristic to be used for this, as we define our framework in general terms and use MDL to identify the structure \emph{type} of the candidates.
	\item[(c)] Given a candidate set ${\cal C}$ of promising subgraphs,
	we show how to mine informative summaries, removing redundancy  by minimizing the cost.
	\end{enumerate}
	
	\method results in a list $M$ of subgraphs, sorted in importance order (= relative cost savings), that together describe the main connectivity of the graph succinctly.
}

In this section we describe the first contribution, the MDL formulation of graph summarization. To enhance readability, we list the most frequently used symbols in Table~\ref{tab:Symbols}.

In general, the Minimum Description Length principle (MDL)~\cite{rissanen:83:integers}, is a practical version of Kolmogorov Complexity~\cite{vitanyi:93:book}, which embraces the slogan {\em Induction by Compression}. 
Given a set of models $\Models$, the best model $\Model \in \Models$ minimizes $\mdl(\Model) + \mdl(\DB\mid\Model)$, in which $\mdl(\Model)$ is the length in bits of the description of $\Model$, and $\mdl(\DB\mid\Model)$ is the length of the description of the data  encoded with $\Model$.  To ensure fair comparison, MDL requires descriptions to be losslesss
%

We consider undirected graphs $G(\mc{V},\mc{E})$ of $n=|\mc{V}|$ nodes, and $m=|\mc{E}|$ edges, without self-loops. Our theory, however, can be easily generalized to graphs in general.
%
 \begin{table}[!bt]
{
 \caption{Description of major symbols.}
 \label{tab:Symbols}
\begin{center}
  \begin{tabular}{lll}
  \toprule
     \textbf{Notation} & \textbf{Description}  \\ \midrule
     $G$ & graph \\
     \A & adjacency matrix of $G$ \\
     $\mc{V}$, $n$ & node-set, \# of nodes of $G$ respectively\\
     $\mc{E}$, $m$ & edge-set, \# of edges of $G$ respectively\\
     fc, nc & full and near clique respectively\\
     fb, nb & full and near bipartite core respectively\\
     st, ch & star, chain resp.  \\
     \multirow{1}{*}{$\Omega$} & vocabulary of structure types\\
      $\mc{C}, \mc{C}_\mi{x}$ & set of all/type $x\in\Omega$ candidate structures \\
     \model & our model, essentially list of structures \\
      $s, t$ & structures in \model \\
     $\mi{area}(s)$ & edges of $G$ (= cells of \A) described by $s$\\
     $|S|$, $|s|$ & cardinality of set $S$, \# of nodes of $s$ respectively\\
     $||s||, ||s||'$ & \# of existing and non-existing edges in $s$ respectively\\
    \M & approximation of \A \xspace deduced by $M$\\
     \E & error matrix, \E \xspace = \M \xspace $\oplus$ \A\\
    $\oplus$ & exclusive OR\\
     $L(G,M)$ & \# of bits to describe $M$, and $G$ using $M$\\
     $L(M)$ & \# of bits to describe model $M$\\
     $L(s)$ & \# of bits to describe structure $s$\\
     \bottomrule
      \end{tabular}
\end{center}
  }
\end{table}
%
%
For the graph summary, we use a set of graph structure types $\Omega$ which we call a \emph{vocabulary}.
Although any graph structure can be a part of the vocabulary,
we choose the 6 most common structures in real-world graphs (\cite{kleinberg99web,PrakashPAKDD2010,TauroPSF01}) that are well-known and understood by the graph mining community: \emph{full} and \emph{near} {cliques} ($\mi{fc},\mi{nc}$), \emph{full} and \emph{near} {bi-partite cores} ($\mi{fb},\mi{nb}$), stars ($\mi{st}$), and chains ($\mi{ch}$).
Compactly, we have $\Omega = \{\mi{fc},\mi{nc}, \mi{fb},\mi{nb},\mi{ch},\mi{st}\}$.
We will formally introduce these types after formalizing our goal.

To use MDL for graph summarization, we need to define what our models $\Models$ are, how a model $\Model \in \Models$ describes data, and how we encode this in bits. We do this next.

\subsection{MDL for Graph Summarization.}

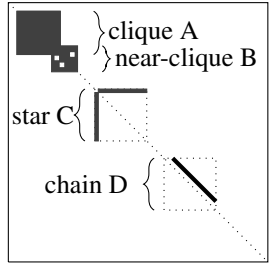
\begin{figure}[tb]
{\centering
\begin{tabular}{c}
	\begin{tikzpicture}[scale=0.23]
		\draw [thin] (0,0) -- (0,15) -- (15,15) -- (15,0) -- (0,0);
		\draw [thin, dotted] (0,15) -- (15,0);

		\filldraw [fill=transparent, black!75] (2.5, 15-2.5) rectangle (4,15-4);
		\filldraw[fill=transparent, black!75] (0.5,15-0.5) rectangle (3,15-3);
      \draw [decorate,decoration={brace,amplitude=3pt},xshift=-2pt,yshift=0pt]
            (5.5,15-2.5) -- (5.5,15-4) node [black,midway,xshift=-0.9cm]
            {};
		\node at (10.3,15-3.30) {near-clique B};

		\filldraw[white] (3,15-3.5) rectangle (3+0.2, 15-3.5-0.2);
		\filldraw[white] (2.7,15-3.1) rectangle (2.7+0.2, 15-3.1-0.2);
		\filldraw[white] (3.6,15-2.9) rectangle (3.6+0.2, 15-2.9-0.2);

		\filldraw[fill=transparent, black!75] (0.5,15-0.5) rectangle (3,15-3);
      \draw [decorate,decoration={brace,amplitude=4pt},xshift=-2pt,yshift=0pt]
            (5,15-0.5) -- (5,15-3) node [black,midway,xshift=-0.9cm]
            {};
		\node at (8.4,15-1.75) {clique A};

		\draw [thin, dotted] (5,15-5) rectangle (8, 15-8);
      \draw [decorate,decoration={brace,amplitude=4pt},xshift=-2pt,yshift=0pt]
            (4.5,15-8) -- (4.5,15-5) node [black,midway,xshift=-0.9cm]
            {};
      \node at (2,15-6.5) {star C};
		\filldraw [fill=transparent, black!75] (5+0.2,15-5) rectangle (8,15-5-0.2);
		\filldraw [fill=transparent, black!75] (5,15-8) rectangle (5+0.2,15-5-0.2);

		\draw [thin, dotted] (9,15-9) rectangle (12, 15-12);
		\draw [ultra thick] (9+0.5,15-9) -- (12, 15-12+0.5);
      \draw [decorate,decoration={brace,amplitude=4pt},xshift=-2pt,yshift=0pt]
            (8.5,15-12) -- (8.5,15-9) node [black,midway,xshift=-0.9cm]
            {chain D};

	\end{tikzpicture}
	\\
\end{tabular}
	\caption{Illustration of our main idea on a toy adjacency matrix:
	\method
	identifies {\em overlapping} sets of nodes, that
	form vocabulary subgraphs (cliques, stars, chains, etc).
	With overlap, \method allows for soft clustering of nodes, as in clique A and near-clique B. Stars look like inverted L shapes (e.g., star C).
	Chains look like lines parallel to the main diagonal (e.g., chain D).
	\label{fig:mainIdea}}
}
\end{figure}

As models ${M}$, we consider ordered lists of graph structures, with possible node (but not edge) overlaps. Each  structure $s \in M$ identifies a patch of the adjacency matrix $\mb{A}$ and describes how it is connected (Fig~\ref{fig:mainIdea}).
%
We refer to this patch, or more formally the edges $(i,j) \in \mb{A}$ that structure $s$ describes, as $\mi{area}(s,M,\mb{A})$, where we omit $M$ and $\mb{A}$ whenever clear from context. 


%
%
Let $\mc{C}_x$ be the set of all possible structures of type $x \in \Omega$, and $\mc{C}$ the union of all of those sets, $\mc{C} = \cup_x{\mc{C}_x}$.  
For example, $\mc{C}_\mi{fc}$ is the set of all possible full cliques. Our model family $\mc{M}$ then consists of all possible permutations of all possible subsets of $\mc{C}$ -- recall that the models $M$ are {\em ordered} lists of graph structures. By MDL, we are after the $M \in \mc{M}$ that best balances the complexity of encoding both $\mb{A}$ and $M$.

Our general approach for transmitting the adjacency matrix is as follows. First, we transmit the model $M$. Then, given $M$, we can build the approximation $\mb{M}$ of the adjacency matrix, as defined by the structures in $M$; we simply iteratively consider each structure $s \in M$, and fill out the connectivity of $\mi{area}(s)$ in $\mb{M}$ accordingly. As $M$ is a summary, it is unlikely that $\mb{M} = \mb{A}$. Still, in order to fairly compare between models, MDL requires an encoding to be lossless. Hence, besides $M$, we also need to transmit the error matrix $\mb{E}$, which encodes the error w.r.t.\ $\mb{A}$.
We obtain $\mb{E}$ by taking the exclusive OR between $\mathbf{M}$ and $\mathbf{A}$, i.e., $\mathbf{E} = \mathbf{M} \oplus \mathbf{A}$. Once the recipient knows $M$ and $\mb{E}$, the full adjacency matrix $\mb{A}$ can be reconstructed without loss.

With this in mind, we have as our main score
$$ L(G,M) = L(M) + L(\mathbf{E}), $$
where
$L(M)$ and $L(\mathbf{E})$ are the numbers of bits that describe the structures, and
the error matrix $\mb{E}$ respectively. The formal definition of the problem we tackle in this paper is:

\vspace{0.5em}
\begin{problem}[Minimum Graph Description Problem]
Given a graph $G$ with adjacency matrix $A$, and the graph structure vocabulary $\Omega$,  
by the MDL principle we are after the smallest model $M$
for which the total encoded length
\[
L(G,M) = L(M) + L(\mb{E})
\]
is minimal, where $\mb{E} = \mathbf{M} \oplus \mathbf{A}$ is the error matrix, and $\mb{M}$ is an approximation of $\mb{A}$ deduced by $M$.
\end{problem}

Next, we formalize the encoding of the model 
and the error matrix.

\subsection{Encoding the Model.}

For the encoded length of a model $M \in \mc{M}$, we have
\vspace{-0.3cm}
\begin{eqnarray}
L(M) &=& \LN(|M|+1) + \LU{|M|+\Omega|-1}{|\Omega|-1} \nonumber\\
&&+ \sum_{s \in M}{\big( -\log \Pr(x(s) \mid M) + L(s) \big)} \nonumber.
\end{eqnarray}
First, we transmit the total number of structures in $M$ using \LN, the MDL optimal encoding for integers $\geq 1$~\cite{rissanen:83:integers}. Next, by an index over a weak number composition, we optimally encode the number of structures of each type $x \in \Omega$ in model $M$. Then, for each structure $s \in M$, we  encode its type $x(s)$ with an optimal prefix code~\cite{cover:06:elements}, and finally its structure. 


To compute the encoded length of a model, we need to define $L(s)$ per structure type:

\noindent\textbf{Cliques.}
For a \emph{full clique}, a set of fully-connected nodes, we first encode the number of nodes, and then their ids:
\vspace{-0.2cm}
$$L(\mi{fc}) = L_\mathbb{N}(|fc|) + \log\binom{n}{|\mi{fc}|}.$$

As $M$ generalizes the graph, we do not require that $\mi{fc}$ \emph{is} a full clique in $G$. If only few edges are missing, it may still be  convenient to describe it as such. Every missing edge, however, adds to the cost of transmitting $\mb{E}$.

As long as they stand out from the background distribution, less dense or \emph{near}-cliques can be as interesting as full-cliques. We encode these as follows:
\vspace{-0.2cm}
\begin{eqnarray}
 L(\mi{nc}) &=& L_\mathbb{N}(|\mi{nc}|) + \log \binom{n}{|\mi{nc}|}  \nonumber\\
  && + \log(|\mi{area}(nc)|) + ||\mi{nc}|| l_1 + ||\mi{nc}||' l_0 \nonumber  .
\end{eqnarray}
\noindent We transmit the number and ids of nodes as above, and edges by optimal prefix codes. We write $||\mi{nc}||$ and $||\mi{nc}||'$ for resp.\ the number of present and missing edges in $\mi{area}(\mi{nc})$. 
Then, $l_1 = -\log ((||\mi{nc}|| / (||\mi{nc}|| + ||\mi{nc}||'))$, and analogue for $l_0$, are the lengths of the optimal prefix codes for resp.\ present and missing edges.
The intuition is that the more dense/sparse a near-clique is, the cheaper  encoding it becomes.
Note that this encoding is exact; no edges are added to $\mb{E}$.

\hide{
\paragraph{Off-Diagonal Rectangles}

Next, we discuss encoding off-diagonal rectangles. These are defined by two, non-empty, non-overlapping node sets $A$ and $B$, with $A,B \subseteq V$.
Like for cliques, we can identify a rectangle as being fully connected, or specify a certain edge density. We refer to these as full off-diagonal rectangles (denoted as $\mi{fr}$) and near off-diagonal rectangles ($\mi{nr}$).

For the encoded length of a full off-diagonal rectangle $\mi{fr}$, we have
\[
L(\mi{fr}) = L_\mathbb{N}(|A|) + \LN(|B|) + \log\binom{n}{|A|, |B|} \quad ,
\]
where we first encode the numbers of nodes in the two sets, and subsequently send their ids in one go by an index over all possible non-overlapping selections of $|A|$ and $|B|$ nodes out of $n$ total. After reading a $\mi{fr}$, we can straightforwardly the relevant $\mi{area}(\mi{fr})$ of $\mb{M}$ with $1$s.

Analogue to above, for the encoded length of a near off-diagonal rectangle $\mi{nr}$ we have
\begin{eqnarray}
  \lefteqn{ L(\mi{nr}) = L_\mathbb{N}(|A|) + \LN(|B|) + \log\binom{n}{|A|, |B|}} \nonumber\\
  & + \log(|\mi{area}(nr)|) + ||\mi{nr}|| l_1 + ||\mi{nr}||' l_0 \nonumber \quad .
\end{eqnarray}
}

\vspace{0.2em}\noindent\textbf{Bipartite Cores.}
Bipartite cores are defined as non-empty, non-intersecting sets of nodes, $A$ and $B$, for which there are edges only \emph{between} the sets $A$ and $B$, and not \emph{within}.

The encoded length of a full bipartite core $\mi{fb}$ is
\vspace{-0.3em}
\[
L(\mi{fb}) = L_\mathbb{N}(|A|) + \LN(|B|) + \log\binom{n}{|A|, |B|}  ,
\]
where we encode the size of $A$, $B$, and then the node ids.

As for cliques, for a near bi-partite cores $\mi{nb}$ we have
\vspace{-0.2cm}
\begin{eqnarray}
  L(\mi{nb}) &=& L_\mathbb{N}(|A|) + \LN(|B|) + \log\binom{n}{|A|, |B|} \nonumber\\
  && + \log(|\mi{area}(nb)|) + ||\mi{nb}|| l_1 + ||\mi{nb}||' l_0 \nonumber .
\end{eqnarray}

\vspace{0.2em}\noindent\textbf{Stars.}
A star is specific case of the bipartite core that consists of a single node (hub) in  $A$ connected to a set $B$ of at least 2 nodes (spokes). 
For $L(\mi{st})$ of a given star $st$ we have
\vspace{-0.2cm}
$$
L(\mi{st}) = L_\mathbb{N}(|\mi{st}|-1) + \log n + \log \binom{n-1}{|\mi{st}|-1},
$$
where
$|\mi{st}|-1$ is the number of spokes. We identify the hub out of $n$ nodes, and the spokes from the remainder.

\vspace{0.2em}\noindent\textbf{Chains.}
A chain is a list of nodes such that every node has an edge to the next node, i.e.\ under the right permutation of nodes, $\mb{A}$ has only the  super-diagonal elements (directly above the diagonal) non-zero.
As such, for the encoded length $L(\mi{ch})$ for a chain $\mi{ch}$ we have
\vspace{-0.4cm}
$$
L(\mi{ch}) = L_\mathbb{N}(|\mi{ch}|-1) + 
\sum_{i=0}^{|\mi{ch}|}{\log (n - i)},
$$
where we first encode the number of nodes in the chain, and then their ids in order. Note $\sum_{i=0}^{|\mi{ch}|}{\log (n - i)} \leq |\mi{ch}|\log n$.


\subsection{Encoding the Error.}
\label{sec:enc:err}

Next, we discuss how we encode the errors made by $\mb{M}$ with regard to $\mb{A}$ 
and store the information in the encoding matrix $\mb{E}$. 
While there are many possible approaches, not all are equally good: we know that the more efficient our encoding is, the less spurious `structure' will be discovered~\cite{miettinen:11:mdl4bmf}.

We hence follow~\cite{miettinen:11:mdl4bmf} and encode $\mb{E}$ in two parts, $\mb{E}^+$ and $\mb{E}^{-}$. The former corresponds to the area of $\mb{A}$ that $M$ does model, and for which $\mb{M}$ includes superfluous edges. Analogue, $\mb{E}^-$ consists of the area of $\mb{A}$ not modeled by $M$, for which $\mb{M}$ lacks edges. We encode these separately as they are likely to have different error distributions. Note that as we know near cliques and near bipartite cores are encoded exactly, we ignore these areas in $\mb{E}^+$.
%
We encode the edges of $\mb{E}^+$, and $\mb{E}^-$ similarly as we do for near-cliques:
\vspace{-0.2cm}
\begin{eqnarray}
L(\mb{E}^+) &= \log(|\mb{E}^+|) + ||\mb{E}^+||l_1 + ||\mb{E}^+||'l_0 \quad \nonumber \\
L(\mb{E}^-) &= \log(|\mb{E}^-|) + ||\mb{E}^-||l_1 + ||\mb{E}^-||'l_0 \quad  \nonumber
\end{eqnarray}
That is, we first encode the number of $1$s in $\mb{E}^+$ (or \ $\mb{E}^-$), after which we transmit the $1$s and $0$s using optimal prefix codes. We choose prefix codes over a binomial as this allows us to efficiently calculate local gain estimates in our algorithm without sacrificing much encoding efficiency.

%
\vspace{0.3em}

\noindent \textbf{Remark.} Clearly, for a graph of $n$ nodes, the search space $\mc{M}$ is enormous, as it consists of {\em all} possible permutations of the collection $\mc{C}$ of {\em all} possible structures over the vocabulary $\Omega$. 
Unfortunately, it does not exhibit trivial structure, such as (weak) (anti)monotonicity, that we could exploit for efficient search. Further, Miettinen and Vreeken~\cite{miettinen:12:tr_mdl4bmf} showed that for a directed graph finding the MDL optimal model of only full-cliques is NP-hard.
Hence, we resort to heuristics.

\vspace{-0.1cm}
\section{VoG: Summarization Algorithm} 
\label{sec:model}
%

Now that we have the arsenal of graph encoding based on the vocabulary of structure types, $\Omega$, 
we move on to the next two key ingredients: finding good candidate structures, i.e., instantiating $\mc{C}$, and then mining informative graph summaries, i.e., finding the best model $M$. 
The pseudocode of \method  
is given in Algorithm \ref{algo:vog}, and the code is available for research purposes at {\small {\small \url{www.cs.cmu.edu/~dkoutra/SRC/VoG.tar}}}$\,$. 

\hide{The pseudocode of our proposed method, \method, 
is given in Algorithm \ref{algo:vog}, while next we give the details of its 3 steps.
%
\begin{algorithm}[t]
\caption{\method}
\label{algo:vog}
\begin{algorithmic}
\STATE \textbf{Input}: graph $\G$,
vocabulary of structure types $\Omega$, subgraphs (output of graph decomposition algorithm).\\
\vspace{0.1cm}
\STATE 1. For each of these subgraphs, use MDL and $\Omega$ to identify the structure type that locally minimizes the encoded cost, and populate the candidate set $\mc{C}$.
\vspace{0.1cm}
\STATE 2. Use the constituent heuristics (\plain, \topten, \tophun, \gnf---see Sec.~\ref{sec:selection}) to select a non-redundant subset from the candidates to instantiate the graph model ${M}$. Pick the model of the heuristic with the lowest description cost.
\vspace{0.1cm}
\RETURN graph summary ${M}$ and its encoding cost.
\end{algorithmic}
\end{algorithm}	
}%

\vspace{-0.2cm}
\subsection{Step 1:~Subgraph Generation.}
\label{sec:graphDecomposition}
Any combination of clustering and community detection algorithms can be used to decompose the graph into subgraphs. These include, but are not limited to 
\textit{Cross-asssociations}~\cite{Chakrabarti:2004}, 
\textit{Subdue}~\cite{cook:94:subdue},
\slashburn \cite{KangF11},
\textit{Eigenspokes} \cite{PrakashPAKDD2010}, and 
\textit{METIS} \cite{Karypis99@METIS}.

\subsection{Step 2:~Subgraph Labeling.}
\label{sec:subgraphLabeling}

Given a subgraph from the set of clusters / communities discovered in the previous step, 
we search for the structure $x \in \Omega$ that best characterizes it, with no or some errors (e.g., perfect clique, or clique with some missing edges, encoded as error).

 \textbf{Step 2.1:~Labeling Perfect Structures.} First, the subgraph is 
tested against our vocabulary structure types for error-free match: full clique, chain, bipartite core, or star. The test for {\it clique} or {\it chain} is based on its degree distribution. A subgraph is {\it bipartite} graph if the magnitudes of its maximum and minimum eigenvalues are equal. To find the node ids in the two node sets, A and B, we use BFS (Breadth First Search) with node coloring.
%
 If one of the node sets has size 1, then the given substructure is encoded as {\it star}.


 \textbf{Step 2.2:~Labeling Approximate Structures.} 
If the subgraph 
does not belong to $\Omega$, 
the search continues for the vocabulary structure type that, in MDL terms, best approximates the subgraph. To this end, we encode the subgraph as each of the 6 candidate vocabulary structures, and choose the structure that has the lowest encoding cost.

Let $m^*$ be the graph model with only one subgraph encoded as structure $\in \Omega$ (e.g., clique) and the additional edges included in the error matrix. For reasons of efficiency, instead of calculating the full cost $L(G,m^*)$ as the encoding cost of each subgraph representation, we estimate the \emph{local} encoding cost $L(m^*) + L(\mb{E}^+_{m^*}) +L(\mb{E}^-_{m^*})$, where $\mb{E}^+_{m^*}$ and $\mb{E}^-_{m^*}$ encode the incorrectly modeled, and unmodeled edges respectively (Sec.~\ref{sec:encoding}). The challenge of the step is to efficiently identify the role of each node in the subgraph (e.g., hub/spoke in a star, member of set $A$ or $B$ in a near-bipartite core, order of nodes in chain) for the MDL representation. We elaborate on each structure next.
%

\noindent \textbf{Clique.}
This is the simplest representation, as all the nodes have the same structural role.
For near-cliques we ignore $\mb{E_\mi{nc}}$, and, so, the encoding cost is $L(\mi{nc})$.

\noindent \textbf{Star.}
The highest-degree node of the subgraph is encoded as the hub, 
and the rest nodes as spokes. 

\noindent \textbf{Bipartite core.}
In this case, the problem of identifying the role of each node reduces to finding the maximum bipartite graph, which is known as max-cut problem, and is NP-hard. 
The need of a scalable graph summarization algorithm makes us resort to approximation algorithms. 
Finding the maximum bipartite graph can be reduced to  semi-supervised classification. We consider  two classes which correspond to the two node sets, $A$ and $B$, of the bipartite graph, and the prior knowledge is that the highest-degree node belongs to $A$, and its neighbors to $B$. To propagate these classes/labels, we employ Fast Belief Propagation (FaBP)~\cite{DBLP:conf/pkdd/KoutraKKCPF11} 
assuming heterophily (i.e., connected nodes belong to different classes). 
For near-bipartite cores $L(\mb{E}^+_{\mi{nb}})$ is omitted.

%
\hide{For the bipartite-core, the unmodeled edges \emph{within} the two sets and incorrectly modeled edges \emph{between} the two sets are stored in $\mb{E_{\mi{fb}}}$. The encoding cost is $L(\mi{fb}) + L(\mb{E}^+_{\mi{fb}}) + L(\mb{E}^-_{\mi{fb}})$. For the near-bipartite core, only the unmodeled edges \emph{within} each of the two node sets are stored in $\mb{E_{\mi{nb}}}$, and the encoding cost is given by $L(\mi{nb}) + L(\mb{E}^+_{\mi{nb}}) + L(\mb{E}^-_{\mi{nb}})$.} 

\noindent \textbf{Chain.}
Representing the subgraph as a chain reduces to finding the longest path in it, which is also NP-hard. Therefore, 
we employ the following heuristic. Initially, we pick a node of the subgraph at random, and find its furthest node using BFS (temporary start). Starting from the latter and by using BFS again, we find the subsequent furthest node (temporary end).
Then we extend the chain by local search. Specifically, we consider the 
subgraph from which 
all the nodes that already belong to the chain, except for its endpoints, are removed. Then, starting from the endpoints we employ again BFS. If new nodes are found during this step, they are added in the chain (rendering it a near-chain with few loops). 
%
The nodes of the subgraph that are not members of this chain are encoded as error in $\mb{E_{\mi{ch}}}$. 

After representing the subgraph as each of the vocabulary structures $x$, we employ MDL to choose the representation with the minimum (local) encoding cost, and add the structure to the candidate set, $\mc{C}$. Finally, we associate the candidate structure with its encoding benefit: the savings in bits for encoding the subgraph by the minimum-cost structure type, instead of leaving its edges unmodeled and including them in the error matrix.

\hide{
\begin{figure}
   \centering
   \includegraphics[width=0.6\columnwidth]{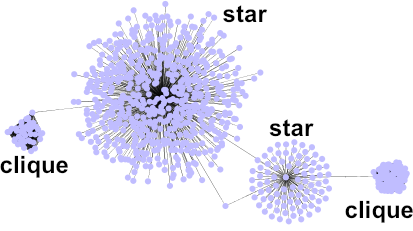}
   \caption{Toy graph for sanity check:
   \method saves 36\% in space,
   (by successfully discovering the 
   two cliques  and two stars that we chained together.)
   }
   \label{fig:synthetic}
\end{figure}
}

\hide{
st 21, 10 11 12 13 14 15 16 17 18 19 20 22 23 24 25 26 27 28 29 30 31 32 33 34 35
fc 44, 45 46 47
fc 43, 44 45 46 47
fc 42 43 44 45 46 47
fc 41 42 43 44 45 46 47
fc 40 41 42 43 44 45 46 47
fc 39 40 41 42 43 44 45 46 47
fc 1 2 3 4 5 6 7 8 9 10
fc 38 39 40 41 42 43 44 45 46 47
fc 37 38 39 40 41 42 43 44 45 46 47
fc 36 37 38 39 40 41 42 43 44 45 46 47
fc 35 36 37 38 39 40 41 42 43 44 45 46 47
}

\begin{algorithm}[t]
\caption{\method}
\label{algo:vog}
{\small
\begin{algorithmic}
\STATE \textbf{Input}: graph $\G$ 
\vspace{0.1cm}
\STATE \textbf{Step 1:~Subgraph Generation.} Generate candidate -- possibly overlapping -- subgraphs using one or more graph decomposition methods.
\STATE \textbf{Step 2:~Subgraph Labeling.} Characterize each subgraph as a perfect structure $x\in\Omega$, or an approximate structure by using MDL to find the type $x$ that locally minimizes the encoding cost. Populate the candidate set $\mc{C}$.
\STATE \textbf{Step 3:~Summary Assembly.} Use the heuristics {\footnotesize \plain, \topten}, {\footnotesize \tophun, \gnf} (Sec.~\ref{sec:selection}) to select a non-redundant subset from the candidate structures to instantiate the graph model ${M}$. Pick the model of the heuristic with the lowest description cost.
\vspace{0.1cm}
\RETURN graph summary ${M}$ and its encoding cost.
\end{algorithmic}
}
\end{algorithm}	


\vspace{-0.2cm}

\subsection{Step 3: Summary Assembly.}
\label{sec:selection}

Given a set of candidate structures, $\mc{C}$, how can we efficiently induce the model $M$ that is the best graph summary?
%
The exact selection algorithm, which considers all the possible ordered combinations of the candidate structures and chooses the one that minimizes
cost, is combinatorial, and cannot be applied to any non-trivial candidate set. Thus, we need heuristics that will give a fast, approximate solution to the description problem. To reduce the search space of all possible permutations, we attach to each candidate structure a quality measure, and consider them in order of decreasing quality. The measure that we use is the encoding benefit of the subgraph, i.e., the number of bits that are gained by encoding the subgraph as structure $x$ instead of noise.
 %
Our constituent heuristics are: 

\noindent $\;$ \emph{\all:} The baseline approach 
gives as graph summary all the candidate structures, i.e., $M = \mc{C}$.

\noindent $\;$ \emph{\topk:} Selects the top $k$ candidate structures, which are sorted in decreasing quality.

\noindent $\;$ \emph{\gnf:} Considers each structure in $\mc{C}$ sequentially, and includes it in $M$: if the graph's encoding cost does not increase, then it keeps the structure in $M$; otherwise it removes it. 
Note that this heuristic is computationally demanding, and works better for small and medium-sized set of candidate structures. 

\enlargethispage{\baselineskip}
 \method employs all the heuristics and picks 
  the best graph summarization, i.e., with the minimum description cost.

\section{Experiments}
\label{sec:exp}
\begin{table*}[ht!]
  \centering
  \caption{
      [Lower is better.] Quantitative analysis of \method with different heuristics: \plain, \topten, \tophun, and \gnf.
      The first column, \original, presents the cost, in bits, of encoding the adjacency matrix with an empty model $M$. For different heuristics we show the relative number of bits needed to describe the adjacency matrix.
      In parentheses, precursored by `u.e.' (for unexplained edges) we give the fraction of edges that are \emph{not} explained by the structures in the model, $M$. The lowest description cost is in bold. 
      }
    \label{tab:encoding_cost}
\footnotesize
  \begin{tabular}{l c@{\hspace{10pt}} r c@{\hspace{10pt}} llll}
\toprule
    \multirow{2}{*}{\textbf{Graph}} && \multirow{2}{*}{\textbf{\original}} && \multicolumn{4}{c}{\textbf{\method}}\\
    \cmidrule{5-8}
    &&   \multicolumn{1}{r}{(bits)} && \textbf{\plain} & \textbf{\topten} & \textbf{\tophun} & \textbf{\gnf}\\
   \midrule
\midrule
\Flickr          && 35\,210\,972      &&  \textbf{81\%} (u.e.: 4\%) & 99\% (u.e.: 72\%) & 97\% (u.e.: 39\%) & 95\% (u.e.: 36\%) \hide{proc 4671 - 33,672,745}             \\
\WWWBarabasi    && 18\,546\,330      &&  \textbf{81\%} (u.e.: 3\%) & 98\% (u.e.: 62\%) & 96\% (u.e.: 51\%) & 85\% (u.e.: 38\%) \hide{proc 4673 - 15,100,727}   \\
\Epinions        && 5\,775\,964       && 82\% (u.e.: 6\%)  & 98\% (u.e.: 65\%) & 95\% (u.e.: 46\%) & \textbf{81\%} (u.e.: 14\%)   \\
\Enron           && 4\,292\,729       && \textbf{75\%} (u.e.: 2\%)  & 98\% (u.e.: 77\%) & 93\% (u.e.: 46\%) & \textbf{75\%} (u.e.: 6\%) \\
\ASOregon       && 475\,912         &&  72\% (u.e.: 4\%)    & 87\% (u.e.: 59\%) & 79\% (u.e.: 25\%) &  \textbf{71\%} (u.e.: 12\%)     \\
\Wikipediachoc  &&  60\,310         &&   96\% (u.e.: 4\%)    & 96\% (u.e.: 70\%) & 93\% (u.e.: 35\%) & \textbf{88\%} (u.e.: 27\%)      \\
\Wikipediacontro   &&  19\,833         &&   98\% (u.e.: 5\%)    & 94\% (u.e.: 51\%) & 96\% (u.e.: 12\%) & \textbf{87\%} (u.e.: 31\%)      \\
\bottomrule
  \end{tabular}
\end{table*}

In this section, we aim to answer the following questions: 
\textbf{Q1.}~ Are the real graphs structured, or 
random and noisy? \\ 
\textbf{Q2.} What structures do the graph summaries consist of, and how can they be used for understanding?\\ \textbf{Q3.}~ Is \method scalable?
%

\begin{savenotes}
\begin{table}[th!]
\centering
\scriptsize
\caption{Summary of graphs used. 
}
\label{tab:datasets}
\setlength{\tabcolsep}{0.13cm}
\footnotesize
\centering
\begin{tabular}{l r@{\hspace{0.2em}}  c@{\hspace{3pt}} rl}
\toprule
   \textbf{Name} & \textbf{Nodes} && \textbf{Edges} & \textbf{Description}  \\
\midrule
     \Flickr \cite{flickr_data} & 404\,733 && 2\,110\,078 & Friendship social network \\
    \WWWBarabasi \cite{snap_web} & 325\,729 && 1\,090\,108 & WWW in nd.edu\\
    \Epinions \cite{snap_web} &  75\,888 &&  405\,740 & Trust graph \\
    \Enron \cite{enron_data} & 80\,163 && 288\,364 &  Enron email \\
    \ASOregon \cite{asoregon_data} & 13\,579 && 37\,448 & Router connections\\
    \Wikipediacontro
    & 1\,005 && 2\,123 & Co-edit graph\\
    \Wikipediachoc
 & 2\,899 && 5\,467 & Co-edit graph\\
\bottomrule
\end{tabular}
\end{table}
\end{savenotes}

The graphs we use in the experiments along with their descriptions are summarized in Table~\ref{tab:datasets}. 
\Wikipediacontro is a co-editor graph on a known Wikipedia controversial topic (name withheld for obvious reasons), where the nodes are users and edges mean that they edited the same sentence.   \Wikipediachoc is a co-editor graph on the `Chocolate' article. The descriptions of the other datasets are given in Table~\ref{tab:datasets}.

\vspace{0.2cm}

\noindent {\em Graph Decomposition.}
In our experiments, 
we use 
\slashburn \cite{KangF11} 
to generate candidate subgraphs, 
because it is scalable, and designed
to handle graphs {\em without} ``cavemen'' structure. Details about the algorithm are given in the Appendix.
We note that \method would only benefit from using the outputs of additional decomposition algorithms.

\subsection{ Q1: Quantitative Analysis }

In this section we apply \method to the real datasets of Table~\ref{tab:datasets}, and evaluate the achieved description cost, and edge coverage, which are indicators of the discovered structures. The evaluation is done in terms of savings w.r.t.~the base encoding (\original) of the adjacency matrix of a graph with an empty model $M$.

Although we refer to the description cost of the summarization techniques, we note that compression itself is \emph{not} our goal, but our \emph{means} for identifying structures important for graph understanding or attention routing. This is also why we do not compare against standard matrix compression techniques. Whereas \method has the goal of describing a graph with intelligible structures, specialized algorithms may exploit any statistical correlations to save bits.


We compare two summarization approaches:
(a) \original: The whole adjacency matrix is encoded as if it contains no structure; that is, $M = \emptyset$, all of $\mb{A}$ is encoded through $L(\mb{E}^-)$; and
(b) \methodsb, our proposed summarization algorithm with the three selection heuristics
(\all, \topten and \tophun, \gnf). For efficiency, for \Flickr and \WWWBarabasi, \gnf considers the top 500 candidate structures. We note that we ignore very small structures; the candidate set $\mc{C}$ includes subgraphs with at least 10 nodes, except for the Wikipedia graphs where the size threshold is set to 3 nodes. Among the summaries obtained by the different heuristics, we choose the one that yields the smallest description length. 
%


Table~\ref{tab:encoding_cost} presents the summarization cost of each technique w.r.t.\ the cost of the \original approach, as well as the fraction of the edges that remains unexplained. The lower the ratios (i.e., the lower the obtained description length), the more structure is identified. For example, \method-\plain describes \Flickr with only 81\% of the bits of the \original approach
, and explains all but 4\% of the edges, which means that 4\% of the edges are not encoded by the structures in $M$. 

\vspace{-0.1cm}
\begin{observation}
Real graphs do have structure; \method, with or w/o structure selection, achieves better compression
than the \original approach that assumes no structure. 
\end{observation}


\gnf finds models $M$ with fewer structures than \all and \tophun, yet generally obtains (much) succinct graph descriptions. This is due to its ability to identify structures that are informative \emph{with regard to what it already knows}. In other words, structures that highly overlap with ones already selected into $M$ will be much less rewarded than structures that explain unexplored parts of the graph.

\subsection{ Q2: Qualitative Analysis } In this section, we showcase how to use \method and interpret its output.

\subsubsection{Graph Summaries}
\label{sec:graph_summaries}

\begin{table*}[tbp]
  \centering
    \caption{ Summarization of graphs by \methodsb (for different heuristics). The most frequent structures are the stars (st) and near-bipartite cores (nb). For each graph and selection technique (heuristic), we provide the frequency of each structure type: `st' for star, `nb' for near-bipartite cores, `fc' for full cliques, `fb' for full bipartite-cores, `ch' for chains, and `nc' for near-cliques. }
    \label{tab:vog_summarization}
\footnotesize  \begin{tabular}{ l c@{\hspace{4pt}} rrrrrr c@{\hspace{4pt}} rr c@{\hspace{4pt}} rrrr c@{\hspace{4pt}} rrrr}
\toprule
 &&		\multicolumn{6}{c}{\textbf{\plain}}  &&
        \multicolumn{2}{c}{\textbf{\topten}}  &&
		\multicolumn{4}{c}{\textbf{\tophun}} &&
		\multicolumn{4}{c}{\textbf{\gnf}}\\
\cmidrule{3-8} \cmidrule{10-11} \cmidrule{13-16} \cmidrule{18-21}
    \textbf{Graph} &&
		\textbf{st} & \textbf{nb} & \textbf{fc} & \textbf{fb} & \textbf{ch} & \textbf{nc}  &&
		\textbf{st} & \textbf{nb} &&
		\textbf{st} & \textbf{nb} & \textbf{fb} & \textbf{ch} && \textbf{st} & \textbf{nb} & \textbf{fc} & \textbf{fb}\\ 
\midrule
\Flickr          && 24\,385 &   3\,750  &    281    &   9     &    -     &  3 &&  10   & -  &&   99   &  1   & -  &  -  &&   415  & - & - & 1\\
\WWWBarabasi    && 10\,027 &   1\,684  &    487    &  120    &    26    &  - &&   9   & 1  &&   83   & 14   & 3  &  -  &&   403  & 7 & - & 16\\
\Epinions        && 5\,204  &   528    &    13     &   -     &    -     &  - &&   9   & 1  &&   99   &  1   & -  &   - && 2\,738 & - & 8 & -\\
\Enron           && 3\,171  &   178    &    3      &  11     &    -     &  - &&  9   & 1  &&  99   &  1   & -  &  -  && 2\,323 & 3 & 3 & 2\\
\ASOregon       &&  489   &    85    &    -      &   4     &    -     &  - &&  10   & -  &&   93   &  6   & 1  &  -  && 399  & - & - & -\\
\Wikipediachoc  &&  170   &    58    &    -      &   -     &    17    &  - &&   9   & 1  &&   87   &  10  & -  &  3  && 101  & - & - & -\\
\Wikipediacontro &&   73   &    21    &    -      &   1     &    22    &  - &&   8   & 2  &&   66   &  17  & 1  &  16 && 35   & - & - & -\\
\bottomrule
  \end{tabular}
\end{table*}

How does \method summarize real graphs? Which are the most frequent structures?
Table~\ref{tab:vog_summarization} shows the summarization results of \method for different structure selection techniques. 
\begin{observation}
The summaries of all the selection heuristics consist mainly of stars, followed by near-bipartite cores. In some graphs, like \Flickr and \WWWBarabasi, there is a significant number of full cliques. 
\end{observation}

%

From Table~\ref{tab:vog_summarization}  
we also observe that \gnf drops uninteresting structures, and reduces the graph summary. Effectively, it filters out the structures that  explain edges already explained by structures in model $M$.  


\subsubsection{Graph Understanding}

Are the `important' structures found by \method semantically meaningful? For sense-making, we analyze the discovered subgraphs in 
the non-anonymized real datasets \Wikipediacontro, and \Enron. 

\begin{figure*}[htbp!]
        \centering
        \begin{subfigure}[b]{0.18\textwidth}
               \centering
               \includegraphics[width=0.95\textwidth]{fig/lcr_whole_top8starsRED}
               \caption{\method: The 8 top ``important'' stars whose centers are denoted with red rectangles.}
               \label{fig:lcrStars2}
        \end{subfigure}
        ~ %
        \begin{subfigure}[b]{0.205\textwidth}
                \centering
                \includegraphics[width=0.95\textwidth]{fig/lcr_whole_top1nb_RED}
                \caption{\method: The most ``important'' bipartite graph (node set $A$ denoted by the circle of red points).}
                \label{fig:lcrTop1nb2}
        \end{subfigure}
        ~
        \begin{subfigure}[b]{0.4\textwidth}
               \centering
               \includegraphics[width=\textwidth]{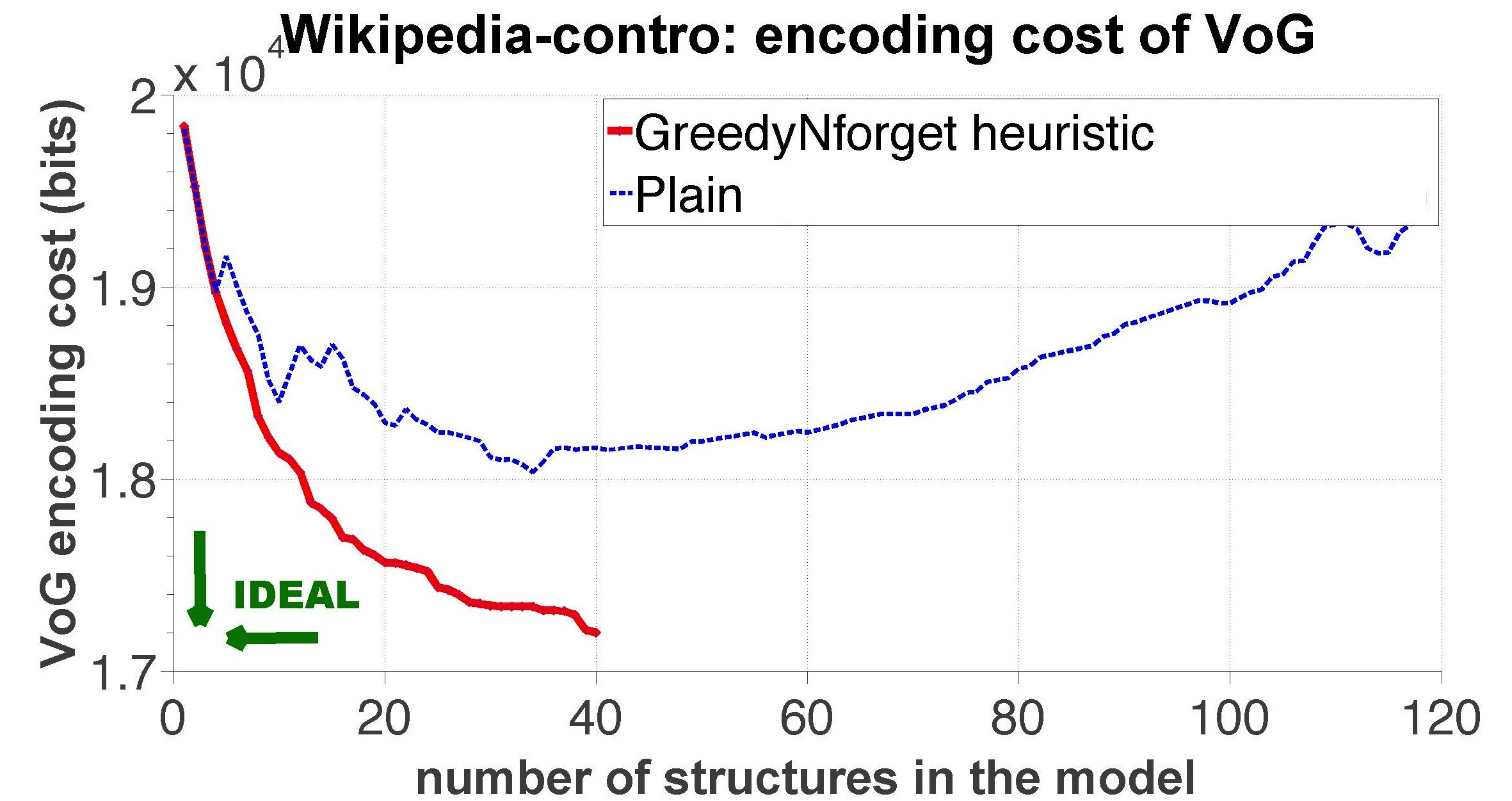}
               \caption{Effectiveness of \gnf (in red). Encoding cost of \method vs.\ number of structures in the model, $M$.}
               \label{fig:lcrCost}
        \end{subfigure}
        \caption{The \method summary of the \Wikipediacontro graph, and effectiveness of the \gnf heuristic. The top 10 structures of the graph summary of \method consist of 8 stars (Fig.~\ref{fig:lcrStars2}) and 2 bipartite graphs (Fig.~\ref{fig:lcrTop1nb2}, \ref{fig:lcrTop2nb}). Fig.~\ref{fig:lcrCost} shows the encoding cost for the \all (dotted blue line) and \gnf (solid red line) heuristics. \gnf minimizes the encoding cost by greedily selecting from a sorted, in decreasing quality order, set of structures the ones that reduce the cost. Thus, \gnf leads to better encoding costs and smaller summaries (here only 40 are chosen) than \all ($\sim$120 structures). }
        \label{fig:lcrMediaWikiCost}
\end{figure*}

\paragraph{Wikipedia--Controversy.} Figs.~\ref{fig:lcrMediaWiki} and~\ref{fig:lcrMediaWikiCost}(a-b) illustrate the original and \method-based visualization of the 
\Wikipediacontro graph.
%
The \method-\topten  summary consists of 8 stars and 2 near-bipartite cores (see also Table~\ref{tab:vog_summarization}).
The 8 star configurations correspond mainly to administrators, such as ``Future\_Perfect\-\_at\_sunrise'', who do many minor edits and revert vandalisms. The most interesting structures \method identifies are the near-bipartite cores, which reflect: (a) the conflict between the two parties, and (b)
an ``edit war'' between vandals and administrators or loyal Wikipedia users.

In Fig.~\ref{fig:lcrCost}, the encoding cost of \method is given as a function of the selected structures. The dotted blue line corresponds to the cost of the \plain encoding, where the structures are added sequentially in the model $M$, in decreasing order of quality (local encoding benefit). The solid red line maps to the cost of the \gnf heuristic. Given that the goal is to summarize the graph in the most succinct way, and at the same time achieve low encoding cost, \gnf is effective.
%



\begin{figure}[htbp!]
        \centering
        ~ %
                \centering
                \includegraphics[width=0.3\columnwidth]{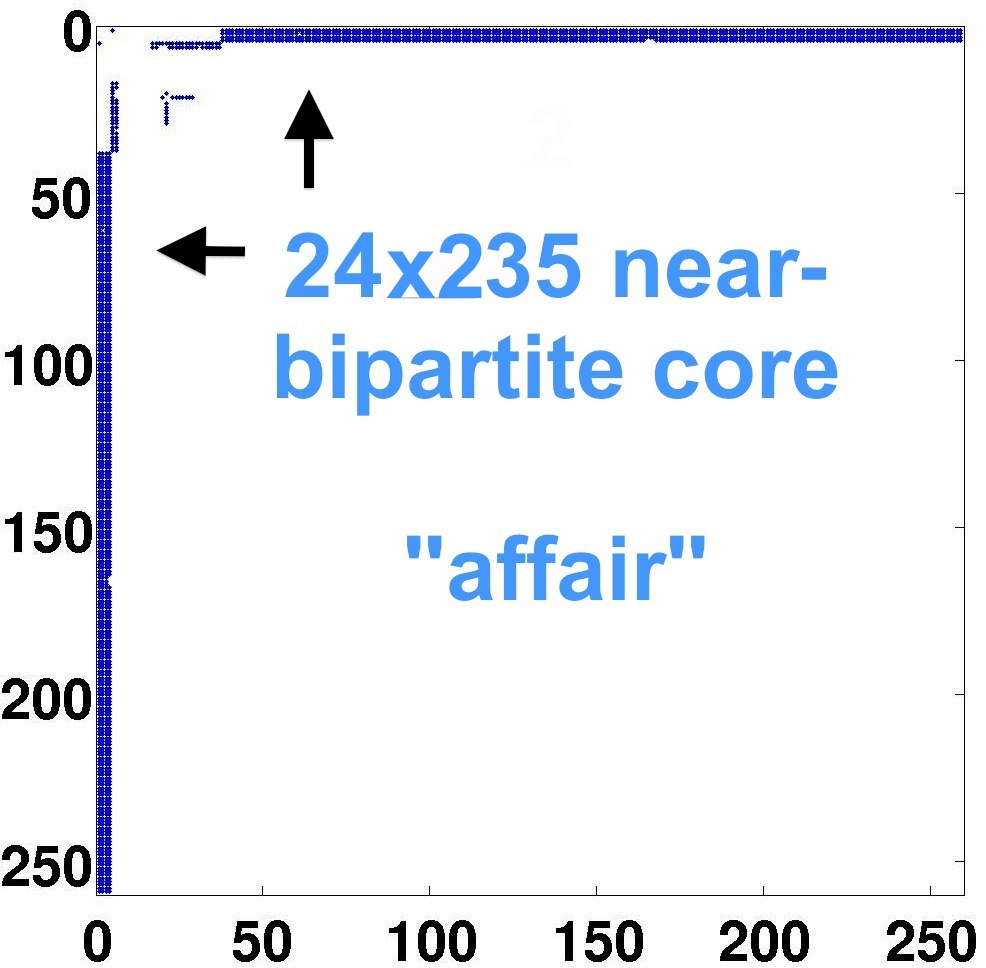}
        \caption{Enron: Adjacency matrix of the top nb core found by \method, corresponding to email communication about an ``affair''.}
                        \label{fig:enron_zoomInMore}
\end{figure}

\paragraph{Enron.}
The \topten summary for Enron has nine stars and one near-bipartite core. The centers of the most informative stars are mainly high ranking officials (e.g., Kenneth Lay with two email accounts, Jeff Skilling, Tracey Kozadinos). 
 As a note, Kenneth Lay was long-time  Enron CEO, while Jeff Skilling had several high-ranking positions in the company, including CEO and managing director of Enron Capital \& Trade Resources.
The near-bipartite core in Fig.~\ref{fig:enron_zoomInMore} is loosely connected to the rest of the graph, and represents the email communication about an extramarital affair, which was broadcast to 235 recipients.

The \method summary for \Wikipediachoc is equally interesting, but not shown here do to lack of space (see the Appendix).

\subsection{ Q3: Scalability of \textbf{\large{\method}} }
\label{sec:scalability}

\begin{figure}[th!]
   \centering
   \includegraphics[width=0.6\columnwidth]{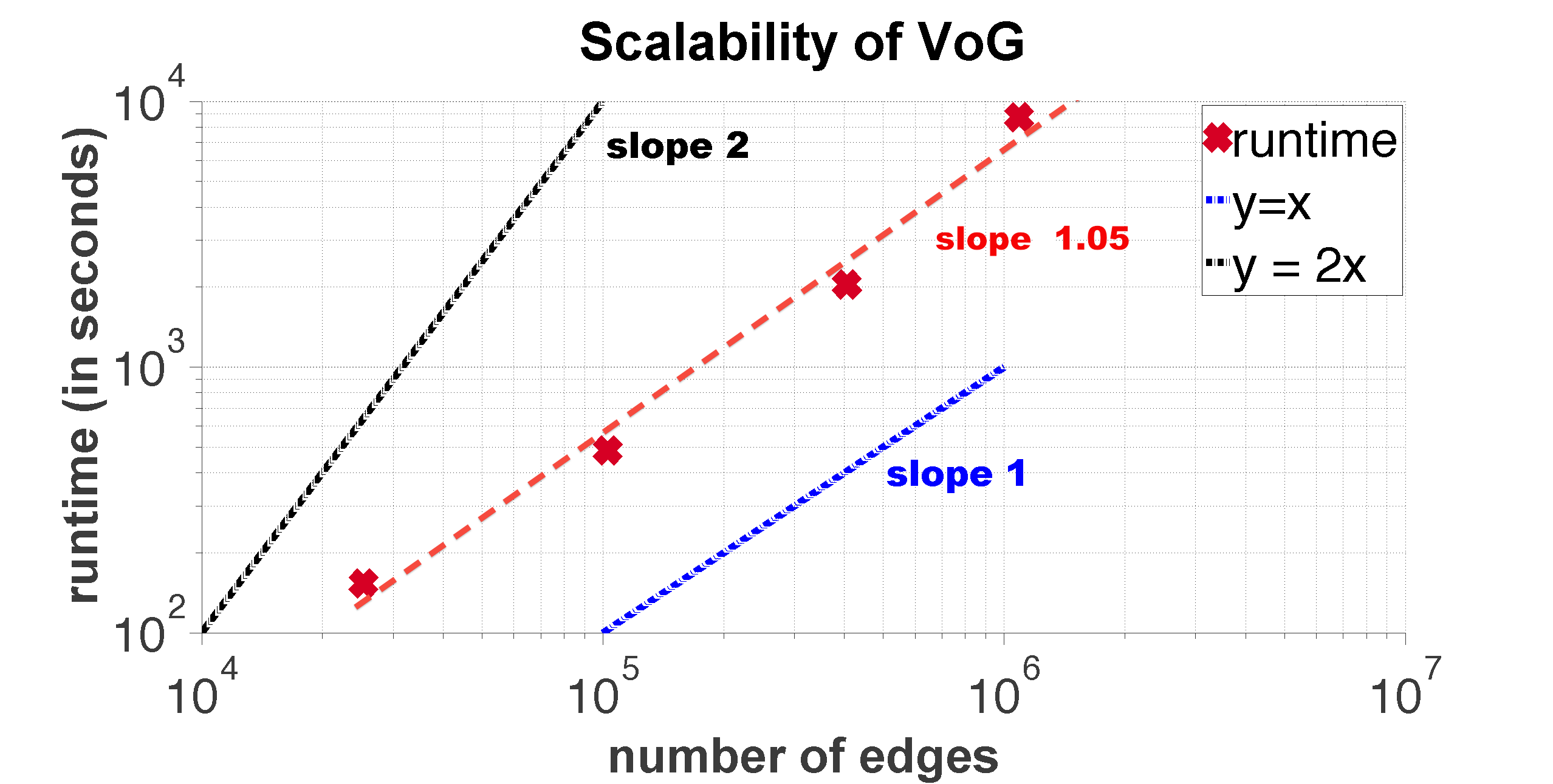}
   \caption{ \method is near-linear on the number of edges. Runtime, in seconds, of \method (\plain) vs.~ number of edges in graph. For reference we show the linear and quadratic slopes.}
   \label{fig:scalability}
\end{figure}

In Fig.~\ref{fig:scalability}, we present the runtime of \methodsb with respect to the number of edges in the input graph.
For this purpose, we induce subgraphs of Notre Dame dataset (\WWWBarabasi) for which we give the dimensions in Table~\ref{tab:datasetsScalability}. We ran the experiments on a Intel(R) Xeon(R) CPU 5160 at 3.00GHz,
with 16GB memory. The structure identification is implemented in Matlab, while the selection process in Python. A discussion about the runtime of \method is also given in the supplementary material.

\begin{observation}
All the steps of \method are 
 designed to be scalable. Fig.~\ref{fig:scalability} shows the complexity is $O(m)$, i.e., \method is near-linear on the number of edges of the input graph.
\end{observation}

\begin{table}[th]
\centering
\footnotesize
\caption{Scalability: Induced subgraphs of \WWWBarabasi.}
\label{tab:datasetsScalability}
\begin{tabular}{lrrl}
\toprule
   \textbf{Name} & \textbf{Nodes} & \textbf{Edges}  \\
\midrule
    \WWWBarabasi-50k  & 49\,780  & 50\,624    \\
	\WWWBarabasi-100k & 99\,854  & 205\,432   \\
    \WWWBarabasi-200k & 200\,155 & 810\,950   \\
    \WWWBarabasi-300k & 325\,729 & 1\,090\,108 \\
\bottomrule
\end{tabular}
\end{table}

\vspace{-0.2cm}

\section{Discussion}
\label{sec:discussion}
 The experiments show that \method successfully 
 solves an important open problem in graph understanding: how to 
 find a succinct summary for a large graph. 
Here we address some questions, that a reader may have. 


\par \noindent
{\bf Q1:}
Why does \method use the chosen vocabulary structures (stars, cliques, etc),
and not other structures? 
\par \noindent
{\bf A1:}
We noticed that these structures appear very often, 
in tens of real graphs,
(e.g.\ in patent citation network,
in phone-call networks, in netflix recommendation system, etc).

\par \noindent
{\bf Q2:} What if a new structure (say, `loops'), proves to be frequent
in real graphs?

\par \noindent
{\bf A2:}
\method can easily be extended to handle new vocabulary terms. 
In fact, MDL will immediately tell us whether a vocabulary set ${\cal V}_1$
is better than a vocabulary set ${\cal V}_2$:
the one that gives best compression wins! 

\par \noindent
{\bf Q3:}
Why not determine automatically which `vocabulary terms' are
best for a given graph?

\par \noindent
{\bf A3:}
Scalability.
Spotting frequent subgraphs 
has the notoriously-expensive subgraph isomorphism problem 
in the inner loop.
Published algorithms on frequent subgraphs (e.g.,~\cite{yan:02:gspan}),
are not applicable here, since they
expect the nodes to have labels 
(e.g., carbon atom, oxygen atom etc.). 

\par \noindent
{\bf Q4:} Why would you focus on compression,
since your goal is pattern discovery and understanding?

\par \noindent
{\bf A4:} Compression is {\em not} our goal; 
it is only our means
to help us find good patterns.
High compression ratios are exactly a sign
that we discovered many redundancies (i.e., patterns) which can be explained in simple terms (i.e., structures), and thus, we understand the input graph better.


\section{Related Work}
\label{sec:related}
Work related to \method comprises the following areas: 

{\bf MDL for Non-Graph Data.}
The MDL principle~\cite{rissanen:83:integers} and compression ~\cite{faloutsos:07:kolmo}
are related to summarization and pattern discovery (
e.g., clustering~\cite{cilibrasi:05:cluster}, 
pattern set mining~\cite{vreeken:11:krimp}).

{\bf Graph Compression.} This includes lexicographic-based compression for web~\cite{BoldiV04}, and social networks ~\cite{ChierichettiKLMPR09}; 
BFS-based techniques~\cite{ApostolicoD09};
multi-position linearizations for neighborhood queries
\cite{MaserratP10};
exploitation of the power-laws in real graphs ~\cite{DBLP:conf/www/LeskovecLDM08} and structural equivalence \cite{ToivonenZHH11};
attribute-based, non-overlapping and covering node grouping \cite{TianHP08,ZhangTP10}.
%
\hide{
Boldi~\cite{BoldiV04} studied the compression of web graphs
using the lexicographic localities;
Chierichetti et al.~\cite{ChierichettiKLMPR09}
extended it to the social networks;
Apostolico et al.~\cite{ApostolicoD09} used BFS for compression.
Maserrat et al.~\cite{MaserratP10} used multi-position linearizations for  neighborhood queries.
\textsc{SlashBurn}~\cite{KangF11} exploits power-law behavior of real world graphs, addressing the `no good cut' problem~\cite{DBLP:conf/www/LeskovecLDM08}.
\cite{TianHP08} presents an attribute-based graph summarization technique with non-overlapping and covering node groups; an automation of this method is given in \cite{ZhangTP10}.
\cite{ToivonenZHH11} uses a node structural equivalence based approach for compressing weighted graphs.
}
None of these works summarize in terms of local structures.
Also, compression is not our goal,
but our \emph{means} to understanding the graph,
by finding informative structures.

\hide{None of the above provide summaries in terms of local structures. Also, we should stress  that we view compression not as the goal,
but as the \emph{means} to understanding  the graph,
by discovering sets of informative structures.}


{\bf Graph Partitioning.}
Assuming away the `no good cut' issue, there are countless graph partitioning algorithms: \textsc{Subdue}~\cite{cook:94:subdue}, 
a frequent subgraph mining algorithm for lossy summaries of attributed graphs; iterative, non-overlapping grouping of nodes with high inter-connectivity~\cite{NavlakhaRS08}; 
MDL-based Boolean matrix factorization for mining full cliques~\cite{miettinen:11:mdl4bmf};
cross-association for near cliques and bi-partite cores.~\cite{chakrabarti:04:xassoc}, or hierarchies~\cite{papadimitriou:08:hierachical}; information-theoretic approaches for community detection \cite{RosvallB06}.

\hide{Assuming away the `no good cut' issue, there are countless graph partitioning algorithms. 
\textsc{Subdue}~\cite{cook:94:subdue}
provides lossy summaries of attributed graphs
based on frequent subgraph mining,
iteratively joining nodes into meta-nodes.
Navlakha et al.~\cite{NavlakhaRS08} follow a similar approach,
by iteratively grouping nodes that see high inter-connectivity.
Their method is hence confined to summarizing a graph in terms
of non-overlapping cliques.
Miettinen and Vreeken~\cite{miettinen:11:mdl4bmf} discuss MDL for Boolean matrix factorization. For directed graphs, such factorizations are summaries in terms of (possibly overlapping) full cliques. 
Chakrabarti et al.~\cite{chakrabarti:04:xassoc}
proposed the cross-association method---hard clustering
 nodes into groups, effectively looking for near-cliques
or near-bi-partite cores.
Papadimitriou et al.~\cite{papadimitriou:08:hierachical} extended
this to hierarchies, again of hard clusters.}

{\bf What sets \method apart:}
None of the above methods
meet all the following specifications (which \method does):
(a) gives a soft clustering, (b) is scalable,
(c) has a large vocabulary of graph primitives
(beyond cliques/cavemen-graphs)
and (d) is parameter-free.


\section{Conclusion}
\label{sec:concl}
We studied the problem of succinctly describing a large graph in terms of connectivity structures. 
Our contributions are:
\vspace{-0.7em}
\begin{itemize}[leftmargin=*,itemindent=0pt,noitemsep] 
  \item \problemFormulation
      We proposed an information theoretic graph summarization technique 
      that uses a carefully chosen vocabulary of graph primitives (Sec.~\ref{sec:encoding}).
  \item \scalability
      We gave \method, an 
      effective method 
      which is 
      near-linear 
      on the number 
       of edges of the input graph  
      (Sec. 
      \ref{sec:scalability}).
  \item \noindent \experiments
      We discussed 
      interesting findings like exchanges between
             Wikipedia vandals and responsible editors on large graphs 
%
      (Sec.~\ref{sec:intro}, \ref{sec:exp}).
\end{itemize}

\hide{
Future work
 includes extending the \method vocabulary to more complex graph structures that we know appear in real graphs, such as core-peripheries,
 (bipartite core whose one set also forms a clique),
 and so-called ``jellyfishes'' (cliques of stars),
 as well as implementing \method in the distributed computing framework like Map-Reduce.
}



\section*{Acknowledgments}
{\small
\thanks{
The authors would like to thank Niki Kittur and Jeffrey Rzeszotarski
for sharing the Wikipedia datasets.
JV is supported by the Cluster of Excellence ``Multimodal Computing and Interaction'' within the Excellence Initiative of the German Federal Government. 
Funding was provided by the U.S. ARO and DARPA under Contract Number W911NF-11-C-0088,
    by DTRA 
   under contract No. HDTRA1-10-1-0120,
   by the National Science Foundation
      under Grant No. IIS-1217559,
   by ARL
   under Cooperative Agreement Number W911NF-09-2-0053,
   and by KAIST under project number G0413002.
   %
   %
   The views and conclusions
   are those
   of the authors and should not be interpreted as representing
   the official policies,
   of the U.S. Government, or other funding parties, and no official endorsement should be inferred.
   The U.S. Government is authorized to reproduce and
   distribute reprints for Government purposes notwithstanding
   any copyright notation here on.
}
}

\vspace{-0.3cm}
\bibliographystyle{abbrv}
\bibliography{bib/ukang,bib/abbrev,bib/bib-jilles,bib/christosref,bib/danaiRef}


\appendix
\section{SlashBurn: Details}


\slashburn is an algorithm for node reordering so that the resulting adjacency matrix has clusters/patches of non-zero elements.
The idea is that removing the top high-degree nodes in real world graphs results in the generation of many small-sized disconnected components (subgraphs), and one giant connected component whose size is significantly smaller compared to the original graph. 
Specifically, \slashburn performs two steps iteratively: 
(a) It removes top high degree nodes from the original graph; 
(b) It reorders the nodes so that the high-degree nodes go to the front, disconnected components to back, and the giant connected component to the middle.
During the next iterations, these steps are performed on the giant connected component.

In this paper, \slashburn is used to decompose graphs, and 
MDL is used to select an appropriate model to encode the subgraphs. 

\hide{
A good node-reordering method will reveal patterns, as well as
large empty areas. 
as shown in
Figure~\ref{fig:slashburn_oregon}
on the Wikipedia-choc network.
}


\section{Toy Example}

To illustrate how \method works, we give an example on a toy graph; we apply \method on the synthetic \Cavemen graph of 841 nodes and 7547 edges, which as shown in Fig.~\ref{fig:synthetic} consists of two cliques separated by two stars. The leftmost and rightmost cliques consist of 42, and 110 nodes respectively; the big star (2nd structure) has 800 nodes, and the small star (3rd structure) 91 nodes.

The \emph{raw} output of the decomposition algorithm (step 1) consists of 
the subgraphs corresponding to the stars, the full left-hand and right-hand cliques, as well as subsets of these nodes. Through MDL, \method first correctly identifies the type of these structures (step 2), and through \gnf it automatically finds the true 4 structures without redundancy (step 3). The corresponding model requires 36\% fewer bits than the `empty' model. We note that one bit gain already corresponds to twice the likelihood. \hide{MDL cost empty: 52665, with structures: 33922}

  \begin{figure}[h]
     \centering
     \includegraphics[width=0.4\columnwidth]{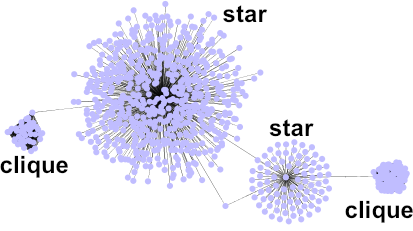}
     \caption{Toy graph for sanity check:
     \method saves 36\% in space,
     by successfully discovering the
     two cliques  and two stars that we chained together.
     }
     \label{fig:synthetic}
  \end{figure}

\section{Time Complexity of \method}
\label{sec:complexity}

For a graph $G(\mc{V},\mc{E})$ of $n=|\mc{V}|$ nodes and $m=|\mc{E}|$ edges, the time complexity of \method depends on the runtime complexity of the
algorithms that compose it, namely the decomposition algorithm, the subgraph labeling,
the encoding scheme $L(G,M)$ of the model, and the structure selection (summary assembly).

For the {\it decomposition} of the graph, we use \slashburn 
which is near-linear on the number of edges of real graphs \cite{KangF11}. 
The {\it subgraph labeling} algorithms in Sec.~4 are carefully designed to be linear in the number of edges of the input subgraph.

When there is no overlap between the structures in $M$, the complexity of calculating the {\it encoding scheme} $L(G,M)$ is $O(m)$. When there is overlap, the complexity is bigger: assume that $s,t$ are two structures $\in M$ with overlap, and $t$ has higher quality than $s$, i.e., $t$ comes before $s$ in the ordered list of structures. Finding how much `new' structure (or area in \A) $s$ explains relative to $t$ costs $O(|M|^2)$. Thus, in the case of overlapping subgraphs, the complexity of computing the encoding scheme is $O(|M|^2 + m)$. As typically $|M| \ll m$, in practice we have $O(m)$.

As far as the {\it selection method} is concerned, the \topk heuristic
that we propose has complexity $O(k)$. 
The \gnf heuristic has runtime $O(|\mc{C}| \times o \times m)$, where $|\mc{C}|$ is the number of structures identified by \method, and $o$ the time complexity of $L(G,M)$. 

\section{Qualitative Analysis of \method}
\label{sec:wikichoc}

In addition to the analysis of \Wikipediacontro and Enron in Sec. 5.2.1, we present qualitative results on the \Wikipediachoc dataset.

\paragraph{Wikipedia--Chocolate.}
The visualization of Wikipedia-choc is similar to Fig. 3 and is omitted.
As shown in Table 4, the \topten summary of \Wikipediachoc contains 9 stars and 1 near-bipartite core. The center of the highest ranked star corresponds to ``Chobot'', a Wikipedia bot that fixes interlanguage links, and thus touches several, possibly unrelated parts of a page. Other stars have as hubs administrators, who do many minor edits, as well as heavy contributors. The near-bipartite core captures the interactions between possible vandals and administrators (or Wikipedia contributors) who were reverting each other's edits resulting to temporary (semi-) protection of the webpage.

\hide{ 
In addition to the analysis of \Wikipediacontro and Enron in Sec.~\ref{sec:graph_summaries}, we present qualitative results on the \Wikipediachoc dataset.

\paragraph{Wikipedia--Chocolate.}
The visualization of Wikipedia-choc is similar to Fig.~\ref{fig:lcrMediaWikiCost} and is omitted.
As shown in Table~\ref{tab:vog_summarization}, the \topten summary of \Wikipediachoc contains 9 stars and 1 near-bipartite core. The center of the highest ranked star corresponds to ``Chobot'', a Wikipedia bot that fixes interlanguage links, and thus touches several, possibly unrelated parts of a page. Other stars have as hubs administrators, who do many minor edits, as well as heavy contributors. The near-bipartite core captures the interactions between possible vandals and administrators (or Wikipedia contributors) who were reverting each other's edits resulting to temporary (semi-) protection of the webpage.
}


\end{document}